\documentclass[aps,prl,twocolumn,superscriptaddress,groupedaddress,nofootinbib]{revtex4-1}  
\usepackage{graphicx}  
\usepackage{dcolumn}   
\usepackage{bm}        
\usepackage{amssymb}   
\usepackage{wasysym}
\usepackage{blindtext}
\usepackage{comment}
\usepackage{color}

\newcommand\blfootnote[1]{
	\begingroup
	\renewcommand\thefootnote{}\footnote{#1}
	\addtocounter{footnote}{-1}
	\endgroup
}

\newcommand{\be}{\begin{equation}}
\newcommand{\ee}{\end{equation}}
\newcommand{\bea}{\begin{eqnarray}}
\newcommand{\eea}{\end{eqnarray}}

\begin{document}

\title{\Large{Direct Higgs-gravity interaction} \\ 
	\Large{and stability of our Universe
}}

\date{\today}

\author{ Vincenzo~Branchina$^{\,a,b\,*}$, Eloisa~Bentivegna$^{\,c\,\dagger}$ Filippo~Contino$^{\,a,b,d\,\ddagger}$
	and Dario~Zappal\`a$^{\,b\,\S}$\vspace{2mm}}

\affiliation{\small $^a$Department of Physics and Astronomy, University of 
	Catania, \\ Via Santa Sofia 64, 95123 Catania, Italy\vspace{1mm}}

\affiliation{\small $^b$INFN, Sezione di Catania, Via Santa Sofia 64,
	95123 Catania, Italy\vspace{1mm}} 

\affiliation{\small $^c$IBM Research UK, The Hartree Centre, Daresbury WA4 4AD, United Kingdom\vspace{1mm}}

\affiliation{\small $^d$Scuola Superiore  di Catania, Via Valdisavoia
	9, 95123 Catania, Italy\vspace{1mm}}

\begin{abstract}
\centerline{\bf ABSTRACT}
\medskip
\noindent
The Higgs effective potential becomes unstable at approximately $10^{11}$ GeV, and if only standard model interactions are considered, the lifetime $\tau$ of the electroweak vacuum turns out to be much larger than the age of the Universe $T_U$. It is well known, however, that $\tau$ is extremely sensitive to the presence of unknown new physics: the latter can enormously lower $\tau$. This poses a serious problem for the stability of our Universe, demanding for a physical mechanism that protects it from a disastrous decay. We have found that there exists a universal stabilizing mechanism that naturally  originates from the nonminimal coupling between gravity and the Higgs boson. As this Higgs-gravity interaction necessarily arises from the quantum dynamics of the Higgs field in a gravitational background, this stabilizing mechanism is certainly present. It is not related to any specific model, being rather {\it natural} and {\it universal} as it comes from fundamental pillars of our physical world: gravity, the Higgs field, the quantum nature of physical laws.
\end{abstract}

\maketitle

\blfootnote{$^*$E-mail: {\tt Branchina@ct.infn.it}\\
	$^\dagger$E-mail: {\tt Eloisa.Bentivegna@ibm.com}\\
	$^\ddagger$E-mail: {\tt filippo.contino@ct.infn.it}\\
	$^\S$E-mail:  {\tt dario.zappala@ct.infn.it} }
\vspace{-0.35cm}

The discovery of the Higgs boson boosted new interest on the stability analysis of the electroweak (EW) vacuum \cite{Cabibbo:1979ay,Flores:1982rv,Lindner:1985uk,Lindner:1988ww,Sher:1988mj,Sher:1993mf,Altarelli:1994rb,Isidori:2001bm,Espinosa:2007qp}, being of crucial importance for our understanding of standard model (SM) and beyond standard model physics and for its impact on cosmological studies, as is the case for Higgs inflation models \cite{Bezrukov:2007ep}.
This renewed interest also prompted a more careful treatment of questions as the gauge invariance of the vacuum decay rate, and the
contribution of zero modes to the quantum fluctuation determinant \cite{Andreassen:2014gha,Andreassen:2017rzq,DiLuzio:2014bua,Endo:2017tsz,Chigusa:2017dux,Andreassen:2014eha}.

It is well known that due to the top loop corrections, the Higgs potential $V(\phi)$ turns over for values of $\phi > v$, where $v \sim 246$ GeV is
the location of the EW minimum, and develops a second minimum at 
$\phi_{\rm min}^{(2)} \gg v$. The location and depth of the latter mainly depend on the Higgs boson and top quark masses, $M_H$ and $M_t$, and for the known values, $M_H \sim 125.09$ GeV and $M_t \sim 173.34$ GeV\,\cite{ATLAS:2014wva,Aad:2015zhl}, it turns out to be much deeper than the EW one, thus being a false vacuum (a metastable state) \cite{Turner,Rees}\footnote{For a different approach to the stabilization problem and renormalization of the effective potential see \cite{Bender:2015uxa,Bender:2012ea,Bender:2013qp}.}. 

To calculate the EW vacuum lifetime $\tau$, that is the tunneling time from the
EW (false) vacuum  to the true one, we need to know the Higgs field dynamics, 
normally described by the (Euclidean) action ($G$ is the Newton constant, $g_{\mu\nu}$ the spacetime metric, $R$ the Ricci scalar),
\be\label{action}
S[\phi, g_{\mu\nu}]=\int d^4 x \sqrt{g} 
\left [-\frac{R}{16 \pi G}+\frac12 g^{\mu \nu} 
\nabla_\mu \phi \, \nabla_\nu \phi + V(\phi)\right ]
\ee
where $V(\phi)$ is the potential to which the Higgs boson is subject. Then we have to seek for the so-called \emph{bounce solutions} to the corresponding (Euclidean) equations of motion\,\cite{Coleman:1977py,Callan:1977pt,Coleman:1980aw}. These are $O(4)$-symmetric solutions that depend only on the radial coordinate $r$, and obey boundary conditions to be specified below. Implementing the $O(4)$ symmetry, the (Euclidean) metric becomes,
\begin{equation}\label{metric}
ds^2= dr^2 + \rho^2(r) d\Omega_3^2\,,
\end{equation}
where $d\Omega_3^2$ is the unit 3-sphere line element,
and $\rho(r)$ is the volume radius of the 3-sphere at 
fixed $r$ coordinate. 

The equations of motion take the form ($\kappa=8\pi G$)\,\cite{Coleman:1980aw},
\begin{equation}\label{gequation}
\ddot \phi + 3 \ \frac{\dot \rho}{\rho} \ \dot \phi 
= \frac{d V}{d \phi} \qquad \dot \rho^2=
1+\frac{\kappa \rho^2}{3} 
\left ( \frac 1 2 \dot \phi^2 - V(\phi) \right ),
\end{equation}
where the first equation is for the Higgs field, while the 
second one is the only Einstein equation left by $O(4)$ 
symmetry. 
The dot indicates derivative with respect to $r$. The boundary conditions 
for the bounce ($\phi_{b}(r)$, $\rho_{b}(r)$) are $\phi_{b}(\infty)=0$ ; $\dot \phi_{b}(0)=0$ ; $\rho_{b}(0)=0$.

\begin{figure*}[t!]
	\centering
	\includegraphics[width=5.4cm,height=3.295cm,angle=360]{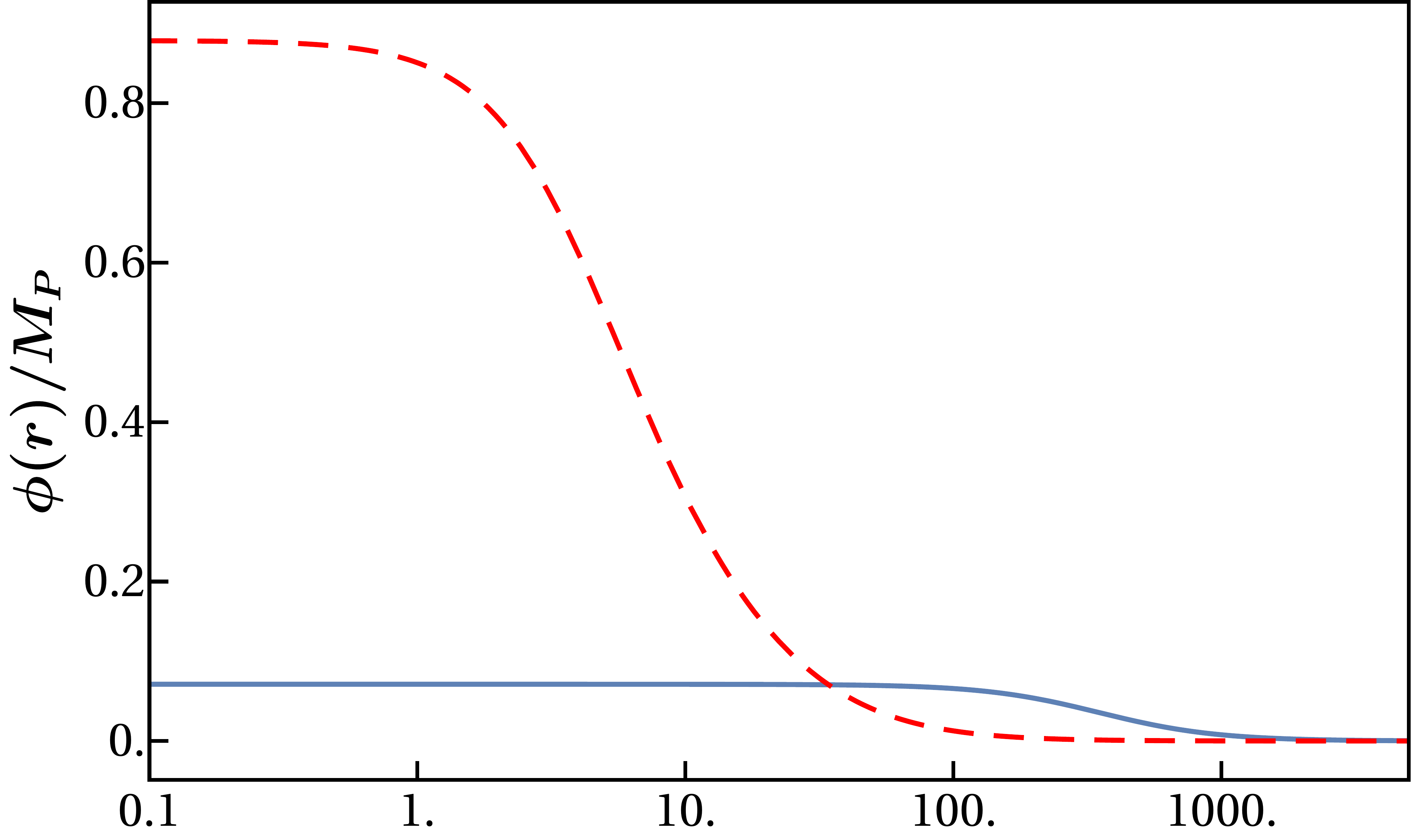}
	\hspace{3mm}
	\includegraphics[width=5.5cm,height=3.295cm,angle=360]{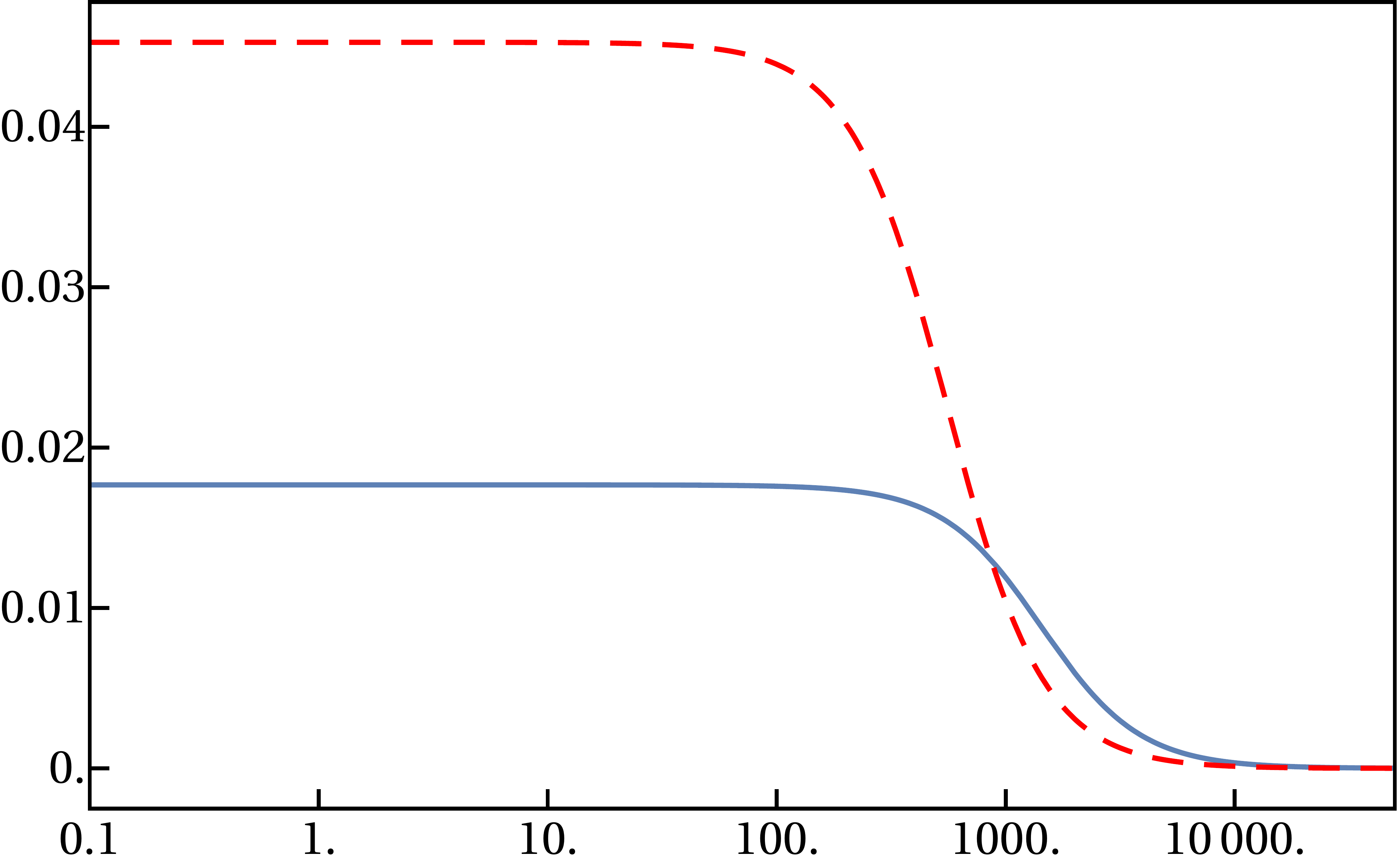}
	\hspace{1mm}
	\includegraphics[width=5.7cm,height=3.295cm,angle=360]{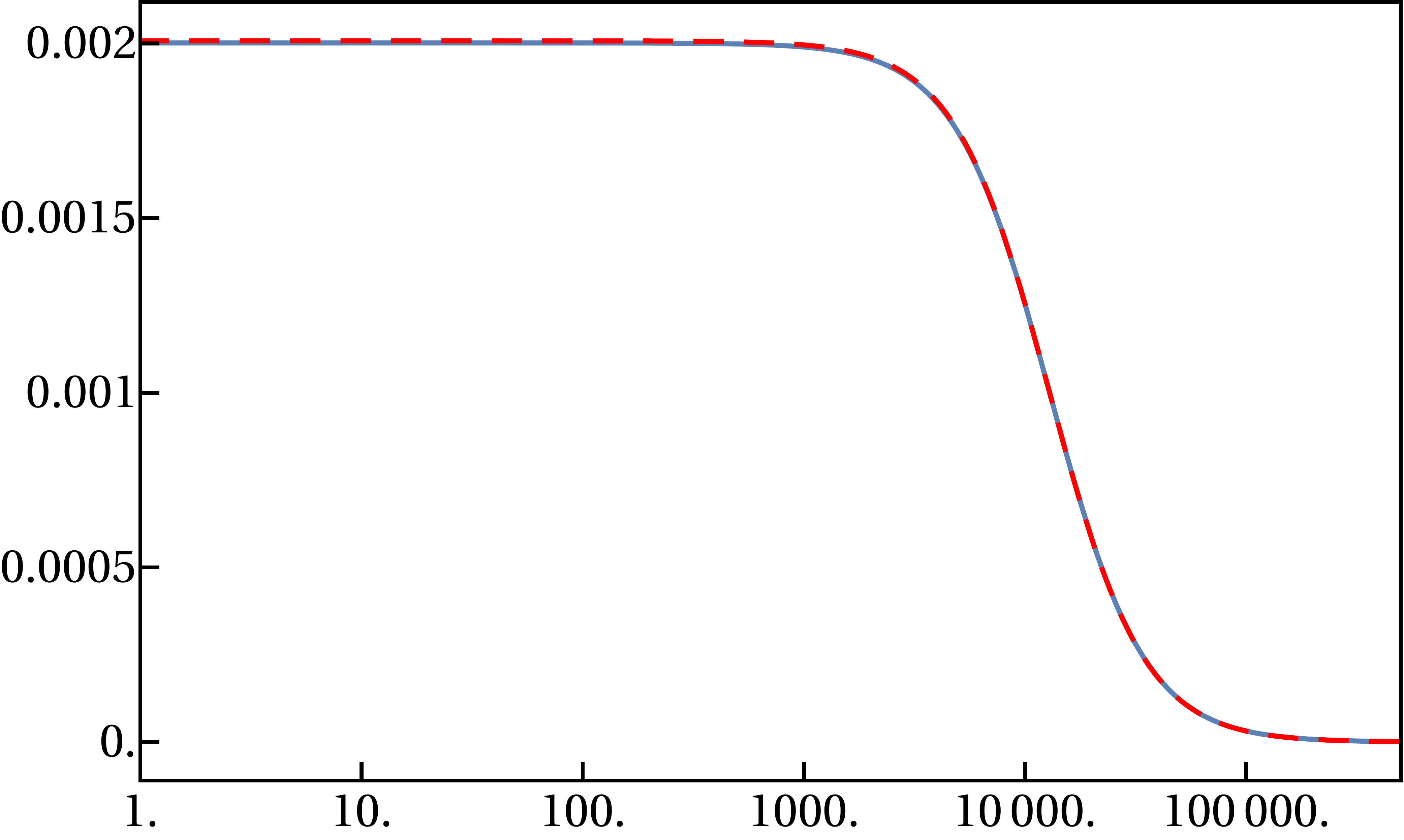}
	\hspace{4mm}
	\includegraphics[width=5.5cm,height=3.295cm,angle=360]{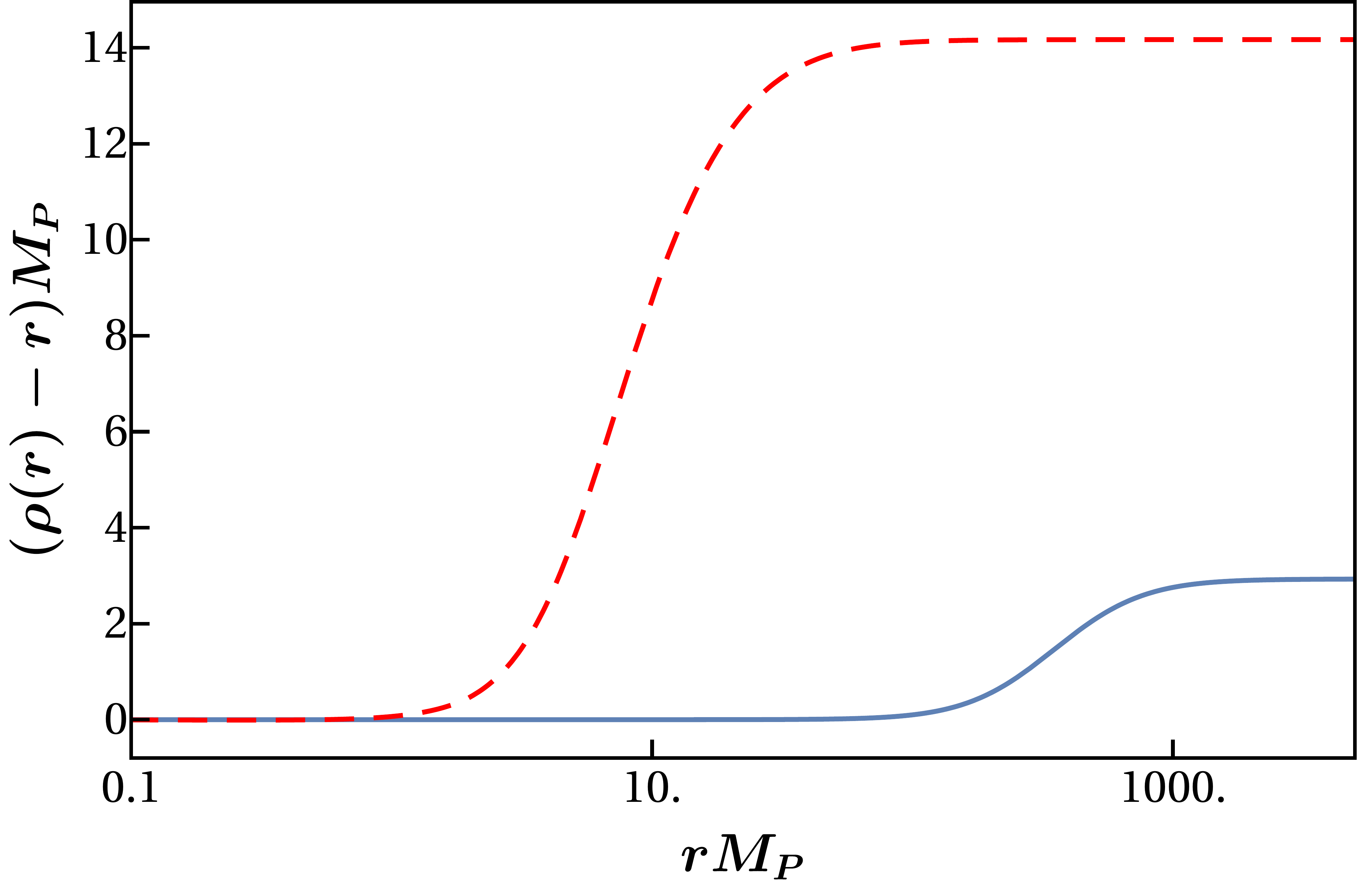}
	\hspace{4mm}
	\includegraphics[width=5.4cm,height=3.295cm,angle=360]{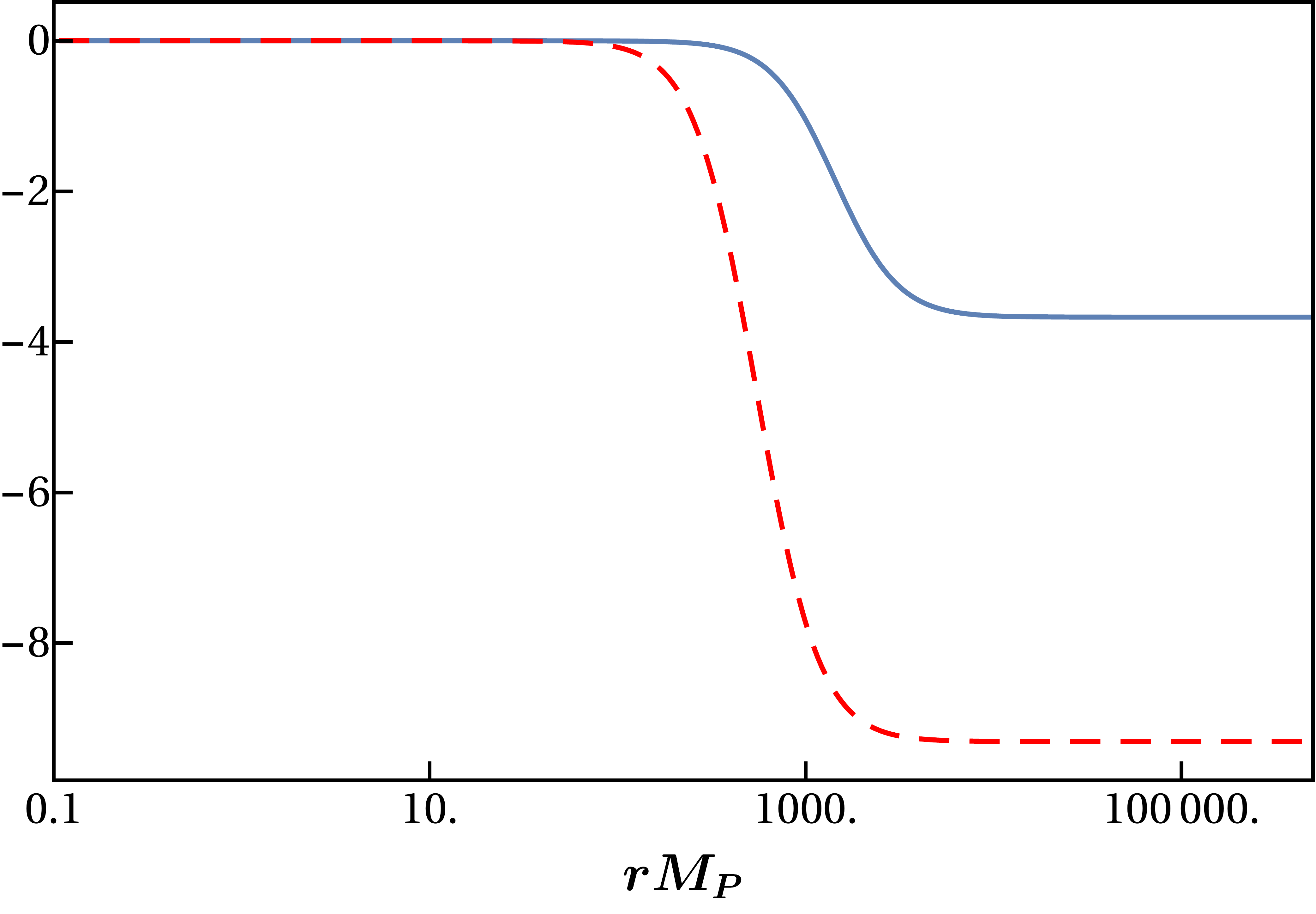}
	\hspace{4mm}
	\includegraphics[width=5.4cm,height=3.295cm,angle=360]{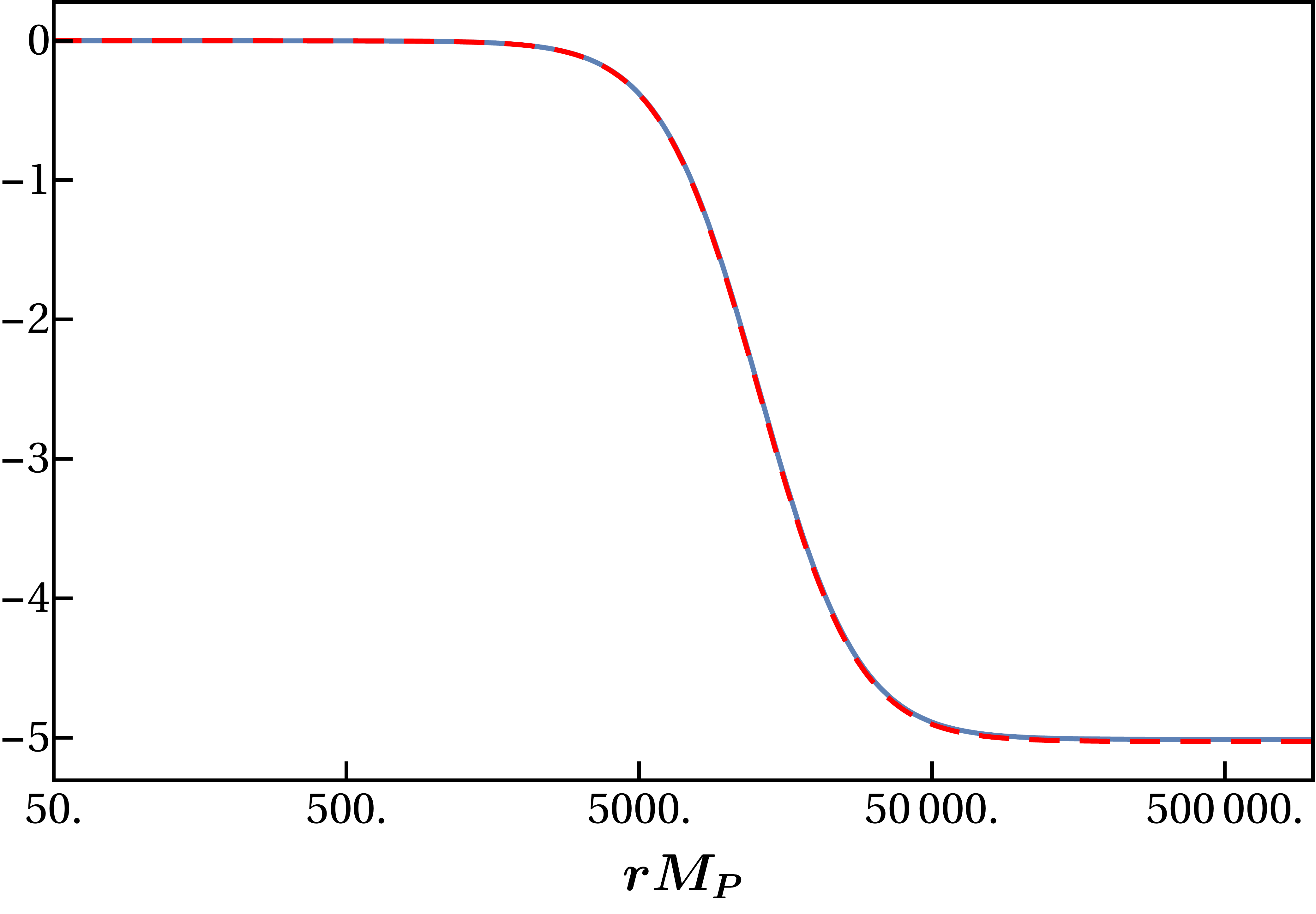}
	\caption{
		{\it Left column. Upper panel}: bounce solution $\phi_{_{\rm NP}}(r)$ (red dashed line) for the action in (\ref{action}), with potential $V(\phi)=V_{\rm eff}(\phi)+V_{_{\rm NP}}(\phi)$, where $\alpha_1=-0.2$ and $\alpha_2=0.125$. The bounce $\phi_{_{\rm SM}}(r)$ (blue solid line) for the SM potential $V_{\rm eff}(\phi)$ alone is also plotted. {\it Lower panel}: the same for $\rho(r)-r$. {\it Middle and right columns}: the same as for the left column for the action with the additional term $\frac12\xi\phi^2 R$ with $\xi=1,10$ respectively.
	}
	\label{fig:1}
\end{figure*} 

The decay rate $\Gamma$ ($= 1/\tau$) from the false to the true vacuum is given by\,\cite{Coleman:1977py,Callan:1977pt,Coleman:1980aw},
\be \label{gamma}
\Gamma =  D\, e^{-[S_{b} - S_{\rm fv}]}\,,
\ee
where $S_{b}\equiv S[\phi_{b},\rho_{b}]$, $S_{\rm fv}$ is the action calculated at the trivial false vacuum solution $(\phi_{\rm fv}, \rho_{\rm fv})$, and $D$ is the quantum fluctuation determinant.

For $O(4)$ symmetric configurations (and in particular for bounces), the action can be written as
\begin{equation}\label{nmcB}
S[\phi,\rho]=-2 \pi^2 \int_0^\infty dr \, \rho^3 V(\phi)\,.
\end{equation}
Moreover, as we take $V(\phi_{\rm fv})=0$, we have $S_{\rm fv}=0$.  

Defining the size $\cal R$ of the bounce as the value of the radial coordinate $r$ such that
$\phi_{b}({\cal R})=\frac 1 2 \phi_{b}(0)$,
the prefactor $D$ in Eq.\,(\ref{gamma}) can be estimated to a good approximation\,\cite{Arnold:1991cv}  as $D \simeq T_U^3 {\cal R}^{-4}$, and $\tau$ then becomes
\be \label{tau}
\tau \simeq \left (\frac{{\cal R}^4}{T_U^3} \right) e^{S_{b}} = \left (\frac{{\cal R}^4}{T_U^4} \right) e^{S_{b}} \, T_U\,.
\ee

In calculating $\tau$, it was usually assumed that $V(\phi)$ can be approximated with the SM Higgs potential. In other words, it was assumed that although high energy (Planckian) NP terms are expected, they can be neglected\,\cite{Isidori:2001bm,Espinosa:2007qp,Degrassi:2012ry,Buttazzo:2013uya}. However, it is now well known that the necessarily present NP terms can have an enormous impact on $\tau$\,\cite{Branchina:2013jra,Branchina:2014rva,Branchina:2014usa,Branchina:2015nda,
Branchina:2016bws,Bentivegna:2017qry}; below we show a specific example [see Eqs.\,(\ref{tauSM}) and (\ref{tauNP})]. 
Before doing that, however, let us consider the SM potential alone.

The SM (renormalization group improved) Higgs potential $V_{\rm eff}(\phi)$ can be approximated as\,\cite{Sher:1988mj,Sher:1993mf,Altarelli:1994rb}:
\be \label{potential}
V_{\rm eff}(\phi) = \frac 1 4 \lambda_{\rm eff} (\phi)\phi^4\,,
\ee
where $\lambda_{\rm eff}(\phi)$ is the quartic running coupling 
$\lambda_{\rm eff}(\mu)$ ($\mu$ is the running scale) with 
$\mu=\phi$\,\cite{Flores:1982rv,Mihaila:2012fm,Chetyrkin:2012rz,Bezrukov:2012sa}.

A good approximation for $V_{\rm eff}(\phi)$ was obtained in\,\cite{Burda:2016mou}, by fitting the two-loop improved Higgs potential with the three parameter function
\be\label{lam}
\lambda_{\rm eff}(\phi)=\lambda_* + \alpha \left ( \ln \frac{\phi}{M_P} 
\right )^2 + \beta \left ( \ln \frac{\phi}{M_P} \right )^4\,,
\ee
where $M_P=1/\sqrt G$ is the Planck mass. The fit gives
\be\label{p}
\lambda_*=-0.013 \quad \alpha=1.4 \times 10^{-5} 
\quad \beta=6.3 \times 10^{-8}\,.
\ee
Taking for $V(\phi)$ the SM potential (\ref{potential}) [with (\ref{lam}) and (\ref{p})], we get \cite{Bentivegna:2017qry}
\begin{equation}\label{tauSM}
\tau_{_{\rm SM}} \sim 10^{661} T_U \,,
\end{equation}
a value much larger than the age of the Universe $T_U$.

New physics at high (Planckian) energies can be parametrized by adding to the SM Higgs potential $V_{\rm eff}(\phi)$ higher powers of $\phi$ as\,\cite{Branchina:2013jra,Branchina:2014rva,Branchina:2014usa,Branchina:2015nda,
Branchina:2016bws,Bentivegna:2017qry,Branchina:2005tu,Branchina:2008pc,Branchina:2018qlf}
\begin{equation} \label{pnp}
V_{_{\rm NP}}(\phi)=\alpha_1 
\frac{\phi^6}{M_P^2}+ \alpha_2 \frac{\phi^8}{M_P^4}\,.
\end{equation}

If we now take $V(\phi)=V_{\rm eff}(\phi)+V_{_{\rm NP}}(\phi)$, and consider for the (dimensionless) couplings $\alpha_1$ and $\alpha_2$ specific values, as for instance $\alpha_1=-0.2$ and $\alpha_2=0.125$, for the EW vacuum lifetime in the presence of NP we find
\begin{equation}\label{tauNP}
\tau_{_{\rm NP}}=10^{-58}\,T_U \,.
\end{equation}

The presence of these NP terms can enormously lower $\tau$\, \cite{Branchina:2013jra,Branchina:2014rva,Branchina:2014usa,Branchina:2015nda,Branchina:2016bws,Bentivegna:2017qry}, to the point that we can get $\tau \ll T_U$. Note that the huge difference between $\tau_{_{\rm SM}}$ and $\tau_{_{\rm NP}}$ is due to a big difference between the bounces in the two cases considered, as can be seen from the left column of Fig.\,\ref{fig:1}.

There must be a mechanism that protects our Universe from a disastrous decay. It has been recently shown that, embedding the SM in supergravity models with discrete R symmetries, a very efficient protective mechanism can be constructed\,\cite{Branchina:2018xdh}. 
In this article we show that
there exists a {\it universal} stabilizing mechanism that arises from the combination of three basic pillars of our physical world: (i) gravity, (ii) the Higgs boson, and (iii) the quantum nature of physical laws. 

In fact, the quantum dynamics of the Higgs field $\phi$ in a gravitational background {\it imposes} a direct interaction between $\phi$ and gravity\,\cite{Callan:1970ze,Birrell:1982ix}
\begin{equation}\label{xi} 
\frac12\xi\phi^2 R\,,
\end{equation}
where $\xi$ is the coupling that measures the strength of this interaction. This term is at the origin of the stabilizing mechanism discussed in this work.

Adding then (\ref{xi}) to (\ref{action}) [and implementing the $O(4)$ symmetry], the equations of motion become
\begin{eqnarray}\label{nmcequation}
&&\ddot \phi + 3 \ \frac{\dot \rho}{\rho} \ \dot \phi=
\frac{d V}{d \phi}+ \xi \phi R \\
&&\label{nmcequation2}
\dot \rho^2=1-\frac \kappa 3 \rho^2 \ 
\frac{-\frac 1 2 \dot \phi^2 + V(\phi) - 
6\xi \frac{\dot \rho}{\rho} \phi \dot \phi}{1-\kappa \xi \phi^2 }\,,
\end{eqnarray}
with boundary conditions as for Eq.\,(\ref{gequation}). 
For $\xi=0$, Eqs.\,(\ref{nmcequation}) and (\ref{nmcequation2}) reduce to Eqs.\,(\ref{gequation}). Moreover, the action for $O(4)$-symmetric configurations takes again the form (\ref{nmcB}).

\begin{table}[t!]
	\renewcommand\arraystretch{1.3}
	\centering 
	\begin{tabular}{|c||cc|}
		
		\hline
		
		$\xi$ \hspace{3mm} & $\tau_{_{\rm SM}}$ 
		\hspace{3mm} & $\tau_{_{\rm NP}}$ \\ \hline \hline
		
		$-15$ \hspace{3mm} & $10^{736}$ \hspace{3mm} 
		& $10^{736}$ \\ \hline

		$-10$ \hspace{3mm} & $10^{726}$ \hspace{3mm} 
		& $10^{726}$ \\ \hline

		$-5$ \hspace{3mm} & $10^{710}$ \hspace{3mm} & 
		$10^{710}$ \\ \hline

		$-1$ \hspace{3mm} & $10^{684}$ \hspace{3mm} & $10^{680}$ \\ \hline
		
		$-0.5$ \hspace{3mm} & $10^{677}$ \hspace{3mm} & $10^{600}$ \\ \hline

		$-0.3$ \hspace{3mm} & $10^{672}$ \hspace{3mm} & $10^{358}$ \\ \hline

		$-0.1$ \hspace{3mm} & $10^{666}$ \hspace{3mm} & $10^{65}$ \\ \hline
		
		$0$ \hspace{3mm} & $10^{661}$ \hspace{3mm} & $10^{-58}$ \\ \hline
		
	\end{tabular}
	\hspace{5mm}
	\begin{tabular}{|c||cc|}
		
		\hline
		
		$\xi$ \hspace{3mm} & $\tau_{_{\rm SM}}$ 
		\hspace{3mm} & $\tau_{_{\rm NP}}$ \\ \hline \hline

		$0.3$ \hspace{3mm} & $10^{660}$ \hspace{3mm} & $10^{-167}$ \\ \hline

		$0.5$ \hspace{3mm} & $10^{668}$ \hspace{3mm} & $10^{23}$ \\ \hline

		$0.7$ \hspace{3mm} & $10^{674}$ \hspace{3mm} & $10^{346}$ \\ \hline
		
		$0.8$ \hspace{3mm} & $10^{676}$ \hspace{3mm} & $10^{512}$ \\ \hline
		
		$1$ \hspace{3mm} & $10^{679}$ \hspace{3mm} & $10^{666}$ \\ \hline
		
		$5$ \hspace{3mm} & $10^{709}$ \hspace{3mm} & 
		$10^{709}$ \\ \hline
		
		$10$ \hspace{3mm} & $10^{725}$ \hspace{3mm} & 
		$10^{725}$ \\ \hline
		
		$15$ \hspace{3mm} & $10^{735}$ \hspace{3mm} & 
		$10^{735}$ \\ \hline
	\end{tabular}
	\caption{Values of $\tau_{_{\rm SM}}$ (second column) and $\tau_{_{\rm NP}}$ (third column) in $T_{U}$ units for different values of $\xi$ (first column). For $\tau_{_{\rm SM}}$, only the SM potential $V_{\rm eff}(\phi)$ is considered. For $\tau_{_{\rm NP}}$, the potential $V_{_{\rm NP}}(\phi)$ of Eq.\,(\ref{pnp}) is added to $V_{\rm eff}(\phi)$, with coupling constants $\alpha_1=-0.2$ and $\alpha_2=0.125$.}
	\label{tab:tab1}
\end{table} 
  
As long as the NP terms are neglected, the inclusion of $\frac12\xi\phi^2 R$ in the action does not change the stability condition of the Universe, as $\tau$ still remains much larger than $T_U$\,\cite{Rajantie:2016hkj}. However, when these terms are taken into account, but the $\frac12 \xi \phi^2 R$ interaction is not included, $\tau$ can be enormously lowered [see Eq.(\ref{tauNP})].

In this article we show that turning on (as we must) the interaction (\ref{xi}), with the exception of a tiny range of values of $\xi$, the EW vacuum lifetime $\tau$ is enormously enhanced and becomes much larger than $T_U$, even in the presence of Planckian NP. This is seen in Table\,\ref{tab:tab1}, where for the Higgs potential we have taken $V(\phi)=V_{\rm eff}(\phi)+V_{_{\rm NP}}(\phi)$, with $\alpha_1=-0.2$, $\alpha_2=0.125$. Table\,\ref{tab:tab1} shows the tunneling time $\tau_{_{\rm NP}}$ (and for comparison $\tau_{_{\rm SM}}$) for different $\xi$. 

\begin{figure}[t!]
	\includegraphics[width=0.4\textwidth]{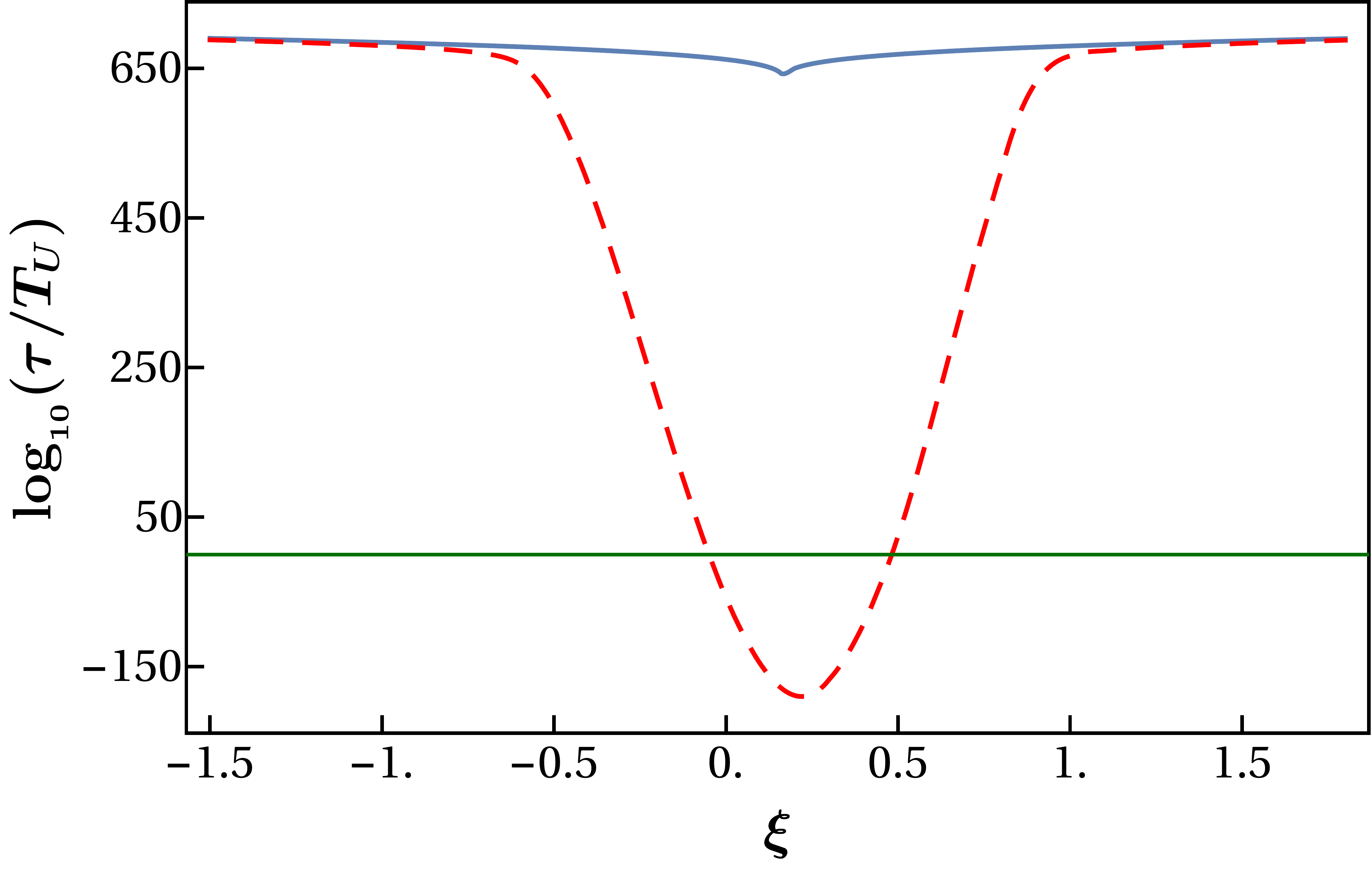}\centering
	\caption{The red dashed line is the $\log_{10}(\tau/T_U)$ as a function of $\xi$ for the Higgs potential $V(\phi)= V_{\rm eff}(\phi) + V_{_{\rm NP}}(\phi)$, where: $\alpha_1=-0.2$ and $\alpha_2=0.125$. The blue line is the $\log_{10}(\tau/T_U)$ for the SM potential $V_{\rm eff}(\phi)$ alone. The green horizontal line separates the region $\tau <T_U$ (lower one) from the region $\tau >T_U$ (upper one). 
	}
	\label{fig:2}
\end{figure}

A graphical representation of the results of Table\,\ref{tab:tab1} is given in Fig.\,\ref{fig:2}, where the decay time $\tau$ [more precisely ${\rm log_{_{10}}} ({\tau}/{T_U})$] as a function of $\xi$ is plotted in the interval $-1.5 \leq \xi \leq 1.8$.   
The range of $\xi$ where $\tau$ is lower than $T_U$ is very tiny ($-0.05 \apprle \xi \apprle 0.5$), and centered around its minimal value $\xi_{\rm min} \sim 0.22$. We observe that, for increasing values of $|\xi|$, $\tau_{_{\rm NP}}$ tends towards $\tau_{_{\rm SM}}$: the interaction $\frac12\xi\phi^2 R$ is so strong to {\it wash out} the destabilizing effect of the NP potential \,(\ref{pnp}). 

The coincidence between $\tau_{_{\rm NP}}$ and $\tau_{_{\rm SM}}$ is due to the fact that with increasing $|\xi|$ the bounces obtained with the Higgs potential $V(\phi)= V_{\rm eff}(\phi)+V_{_{\rm NP}}(\phi)$ tend towards the SM ones, as can be seen from Fig.\,\ref{fig:1}. In fact, actually $\phi_{_{\rm SM}}(0)$ and $\phi_{_{\rm NP}}(0)$ both decrease with increasing $\xi$, and reach the value $\phi(0)\sim 0.002$ for $\xi=10$. For further increasing values of $\xi$, not presented in the figure, $\phi_{_{\rm SM}}(0)$ and $\phi_{_{\rm NP}}(0)$ still coincide and take lower and lower values. For negative $\xi$, the same trend is observed for increasing $|\xi|$.

Now we estimate (for these sufficiently large values of $|\xi|$) the relative weight in the equations of motion (\ref{nmcequation}) and (\ref{nmcequation2}) of the two terms  $\phi^4$ and $\phi^6$ in the potential $V(\phi)=V_{\rm eff}(\phi)+V_{_{\rm NP}}(\phi)$ by considering the ratio
\be\label{rapporto}
A(\phi)=\frac{\alpha_1 \phi^6}{(\lambda/4)\phi^4} = \frac{4 \alpha_1}{\lambda}\phi^2.
\ee

Being $\phi(0) = {\rm max }\,\phi_{b}(r)$ and  $\phi(0) \ll 1$, we find $A(\phi) \ll 1$ (Planck units), so that the (potentially  destabilizing) $\phi^6$ term is very much suppressed as compared to 
the standard $\phi^4$ term. It is then not surprising that the bounce solution for the potential $V_{{\rm eff}}(\phi) + V_{_{\rm NP}}(\phi)$ converges to the corresponding bounce for  $V_{{\rm eff}}(\phi)$ alone.

\begin{figure*}[t!]
	\centering
	\includegraphics[width=5.5cm,height=4.5cm,angle=360]{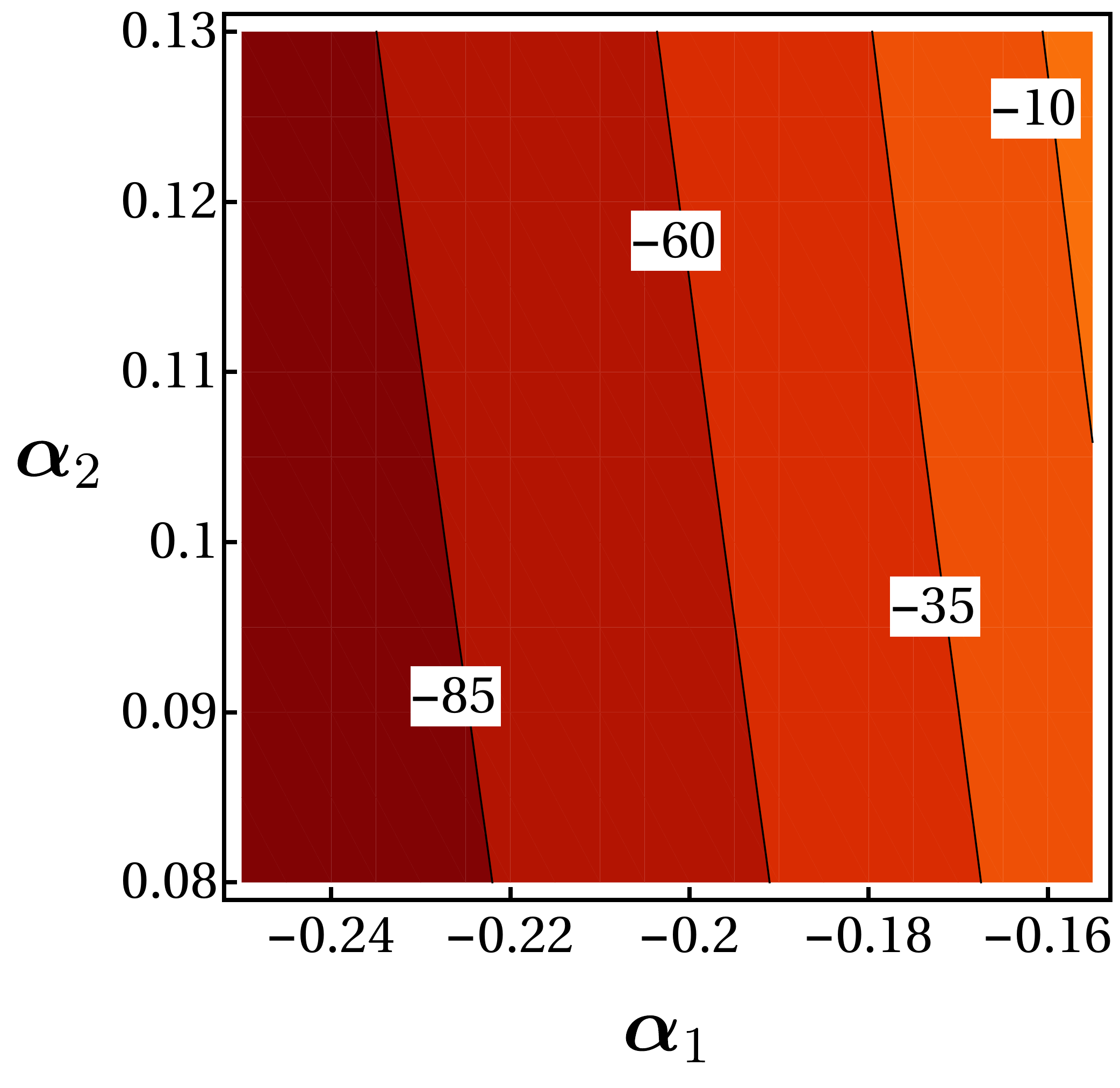}
	\hspace{10mm}
	\includegraphics[width=5cm,height=4.5cm,angle=360]{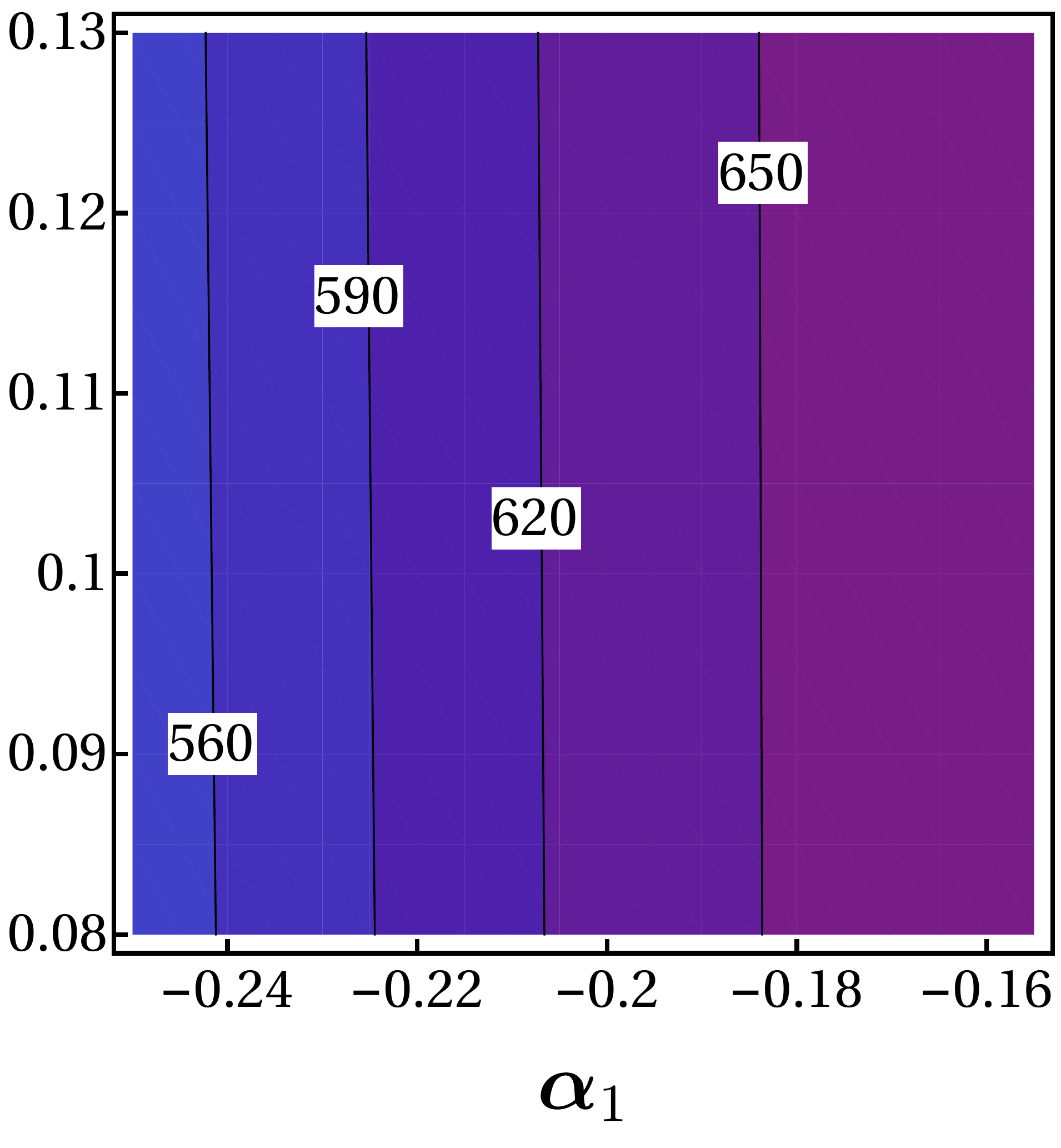}
	\caption{{\it Left panel}. Stability diagram in the $(\alpha_1, \alpha_2)$ plane for the range $-0.25 \leq \alpha_1 \leq -0.16$ , $0.08 \leq \alpha_2 \leq 0.13$,  when $\xi=0$. {\it Right panel}. Stability diagram for $\xi=0.9$ in the same region of the $(\alpha_1, \alpha_2)$ plane.
	}
	\label{fig:3}
\end{figure*} 
\begin{figure*}[t!]
	\centering
	\includegraphics[width=5.5cm,height=4.5cm,angle=360]{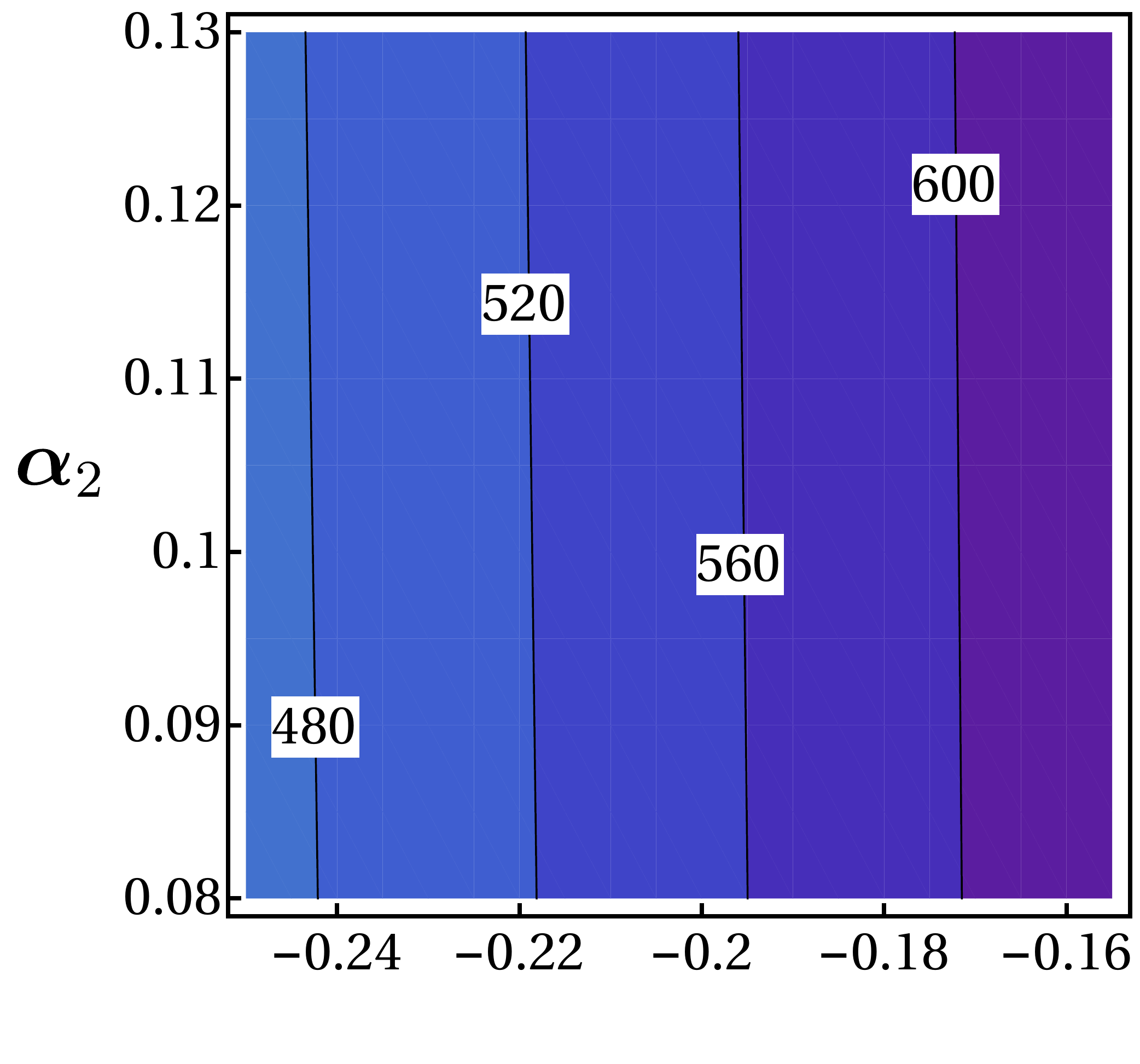}
	\hspace{10mm}
	\includegraphics[width=5cm,height=4.5cm,angle=360]{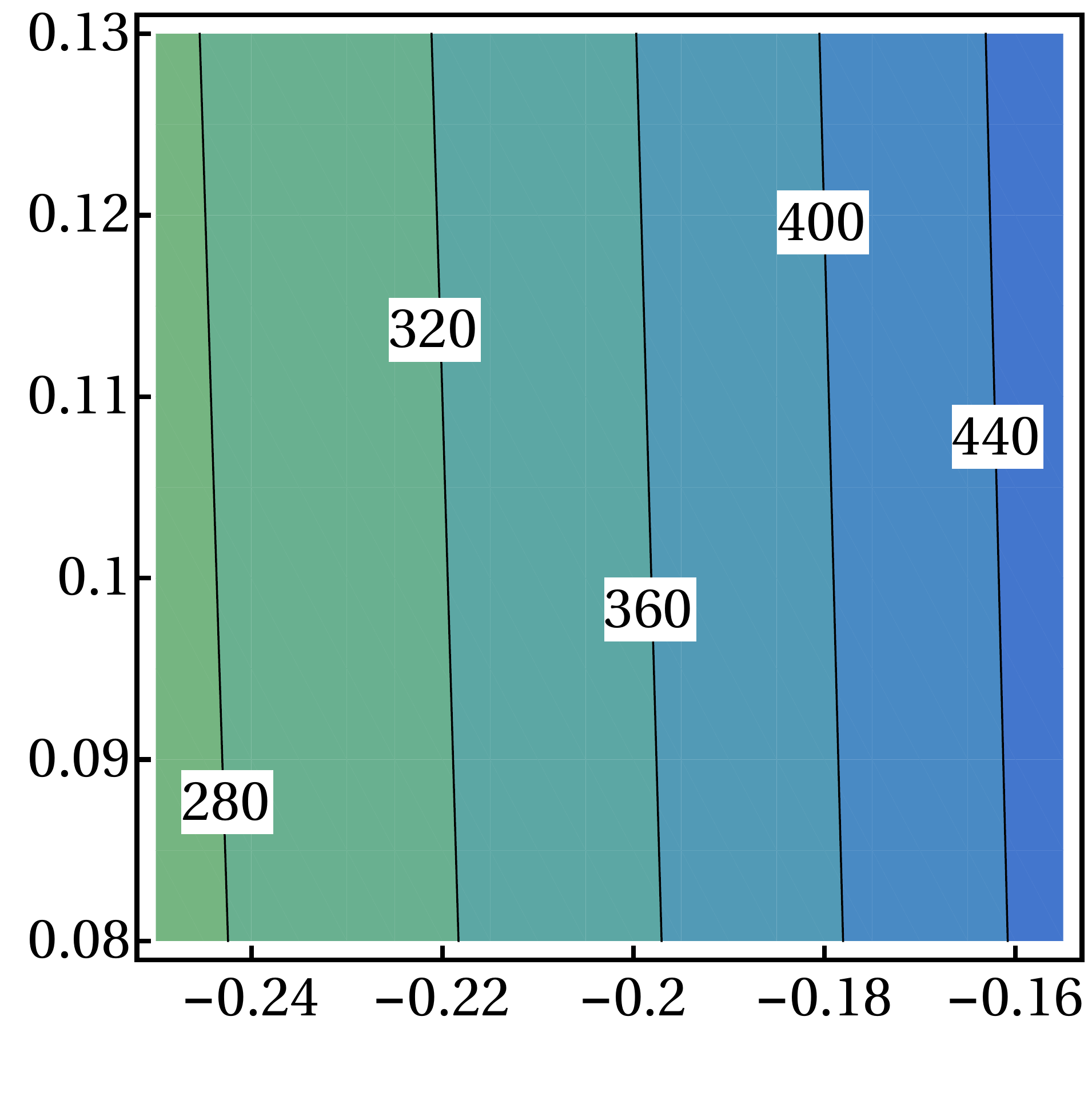}
	\hspace{10mm}
	\includegraphics[width=5.5cm,height=4.5cm,angle=360]{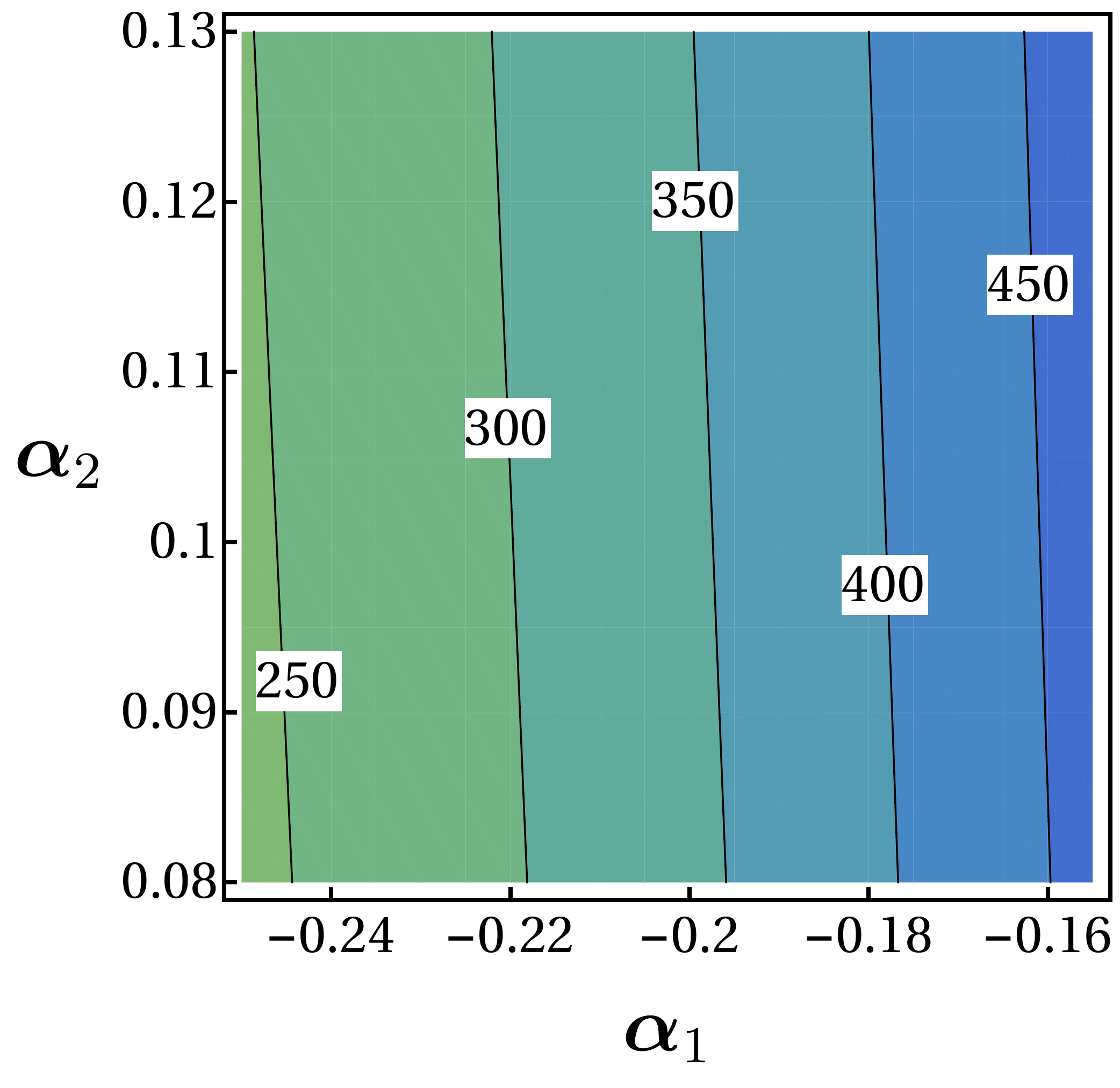}
	\hspace{10mm}
	\includegraphics[width=5cm,height=4.5cm,angle=360]{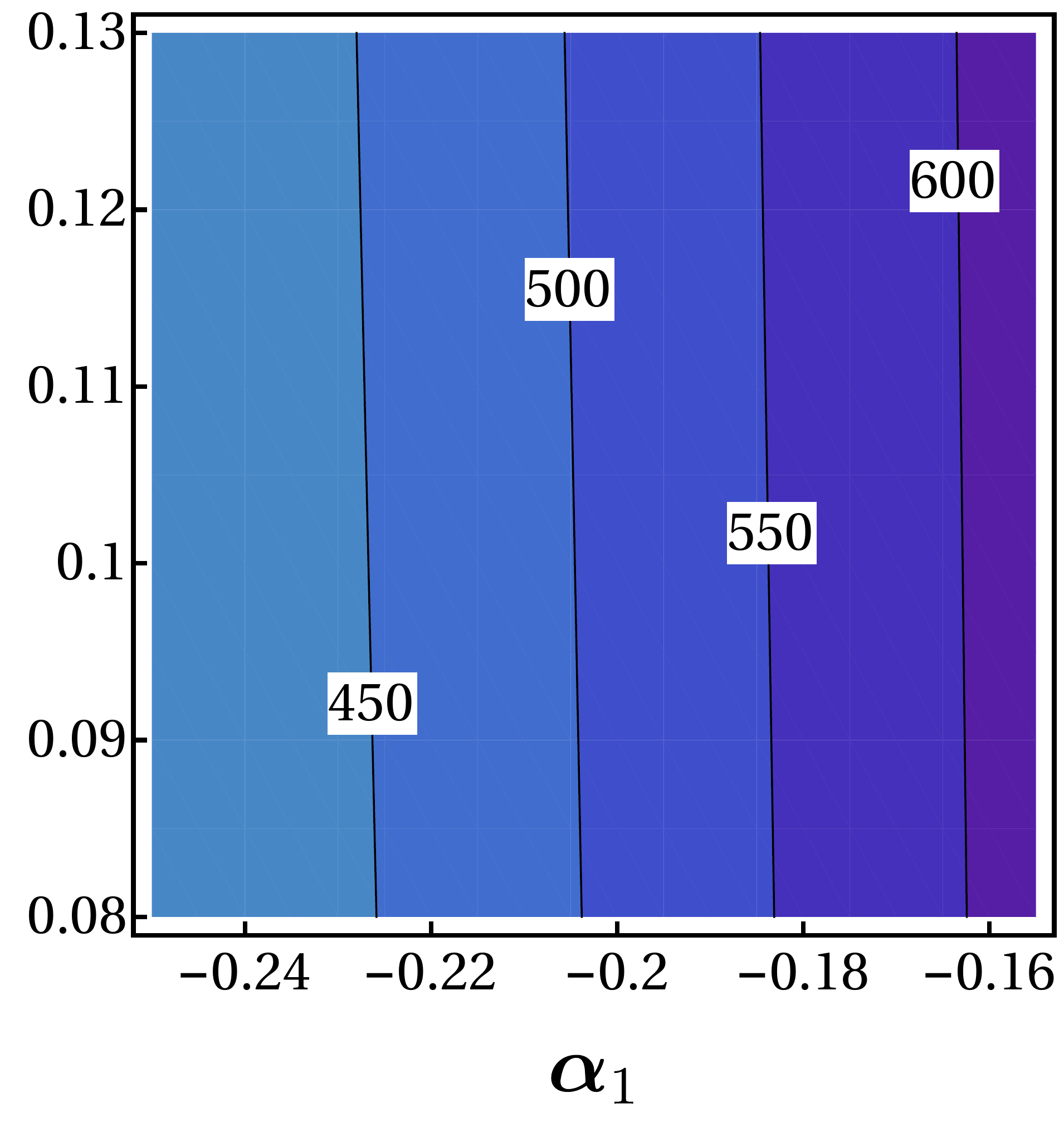}
	\caption{{ Stability diagrams in the ($\alpha_1, \alpha_2$) plane for the potential $V(\phi)=V_{\rm eff}(\phi)+V_{_{\rm NP}}(\phi)$, with $\alpha_1$ and $\alpha_2$ in the same ranges as in Fig.\,\ref{fig:3}. From left to right, from top to bottom  $\xi=-0.4, -0.3,\, 0.7,\, 0.8$. The first two values of $\xi$ are on the left of $\xi_{\rm min}$ (the value of $\xi$ where $\tau$ reaches its minimal value), the last two ones on the right side.  
		}
	}
	\label{fig:4}
\end{figure*}

Finally we see why $\tau_{_{\rm NP}}$ and $\tau_{_{\rm SM}}$ coincide. From (\ref{nmcB}) we see that $S_b$ at the bounce $(\phi_{_{\rm NP}}(r), \rho_{_{\rm NP}}(r))$ is
\be \label{Bsplit}
S_{_{\rm NP}} = -2 \pi^2 \int_0^\infty dr \, \rho_{_{\rm NP}}^3 \Big [ V_{\rm eff}(\phi_{_{\rm NP}}) + V_{_{\rm NP}} (\phi_{_{\rm NP}}) \Big ]\,.
\ee
As for increasing $|\xi| $ we have $(\phi_{_{\rm NP}}(r), \rho_{_{\rm NP}}(r)) \to (\phi_{_{\rm SM}}(r), \rho_{_{\rm SM}}(r))$, Eq.\,(\ref{Bsplit}) can be replaced with
\be \label{Bsplitty}
S_{_{\rm NP}} =   -2 \pi^2 \int_0^\infty dr \, \rho_{_{\rm SM}}^3 \Big [ V_{\rm eff}(\phi_{_{\rm SM}}) + V_{_{\rm NP}} (\phi_{_{\rm SM}}) \Big ]\,.
\ee
For the argument given above, the second term in the rhs of Eq.\,(\ref{Bsplitty}) is negligible as compared to the first one, and having $\phi_{_{\rm SM}}(r)$ and $\phi_{_{\rm NP}}(r)$ practically the same size $\mathcal R$, it follows that $\tau_{_{\rm NP}}$ and $\tau_{_{\rm SM}}$ coincide. 

The enormous stabilizing effect of the Higgs-gravity interaction can be further illustrated by comparing values of $\tau$ calculated at different values of $\xi$ (e.g. $\xi=0$, $\xi=0.9$) in a region of the parameter space $(\alpha_1, \alpha_2)$ where in the $\xi=0$ case $\tau$ is always lower than $T_U$. For $\alpha_1$ and $\alpha_2$ we chose the ranges $-0.25 \leq \alpha_1 \leq -0.16$, $0.08 \leq \alpha_2 \leq 0.13$. 
Figure\,\ref{fig:3} shows the results. The left panel is the stability diagram for the $\xi=0$ case, the right one for $\xi=0.9$. The black lines are level curves with the same value of $\tau$, and the numbers on the top of them are ${\rm log_{_{10}}}(\tau/T_U)$. The red color scale of the left panel, ranging from darker to lighter (left to right), indicates increasing values of $\tau$; as said above, $\tau < T_U$ in the whole region. The right panel is the stability diagram for $\xi=0.9$.
The blue color scale again indicates increasing values $\tau$ going from left to right. The values of $\tau$ have enormously increased, and in the same region of the $(\alpha_1, \alpha_2)$ plane they turn out to be much larger than $T_U$. The destabilizing effect of the NP terms is entirely {\it washed out} by the direct coupling between the Higgs field and gravity. In Fig.\,\ref{fig:4} we consider other values of $\xi$ ($\xi=-0.4,-0.3,0.7,0.8$) that confirm these results. 

The lesson is clear. If we do not take into account the direct Higgs-gravity interaction, NP terms can strongly destabilize the EW vacuum, and without a knowledge of high energy new physics, in particular without a complete theory of quantum gravity, we cannot draw any conclusion on the ultimate fate of our Universe. The Higgs-gravity interaction term, whose presence is guaranteed by exceptionally well-known experimental facts (gravity, the Higgs boson, and the quantum nature of physical laws), acts as a {\it universal stabilizing mechanism}, that washes out any potentially destabilizing effect from high energy new physics (for instance from unknown quantum gravity), protecting our universe from a disastrous decay.  

\paragraph*{Acknowledgments}
This work is carried out within the INFN project QFT-HEP and is supported in part by the Polish National Science Centre HARMONIA Grant No. UMO-2015/18/M/ST2/00518 (2016--2019).

\end{document}